\documentclass[prb,aps,preprint,superscriptaddress,11pt]{revtex4-1}
\usepackage{graphicx}
\usepackage{graphicx}
\usepackage{dcolumn}
\usepackage{bm}
\usepackage{graphicx}
\usepackage{dcolumn}
\usepackage{threeparttable}
\usepackage{gensymb}
\usepackage[english]{babel}
\usepackage{inputenc}  
\usepackage{physics}
\usepackage{graphicx}
\usepackage{dcolumn}
\usepackage{bm}
\usepackage{graphicx}
\usepackage{dcolumn}
\usepackage{threeparttable}
\usepackage{gensymb}
\usepackage[english]{babel}
\usepackage{inputenc}  
\usepackage{physics}
\usepackage{bm} 
\usepackage{amsmath}  
\usepackage{amsfonts} 
\usepackage{bm} 
\usepackage{amsmath}  
\usepackage{amsfonts} 
\begin{document}
\title[]{Resonance fluorescence from an atomic-quantum-memory compatible single photon source based on GaAs droplet quantum dots}
\author{Laxmi Narayan Tripathi}
\affiliation{Technische Physik and Wilhelm-Conrad-R\"ontgen Research Center for Complex Material Systems, Universit\"at W\"urzburg, Am Hubland, W\"urzburg-97074,	Germany}
\affiliation{Department of Physics, Birla Institute of Technology, Mesra, Ranchi 835215, Jharkhand, India}
 \author{Yu-Ming He}
 \affiliation{ National Laboratory for Physical Sciences at Microscale and Department of Modern Physics, University of Science and Technology of China, Shanghai
 	Branch, Shanghai 201315, China}
 \affiliation{ CAS-Alibaba Quantum Computing Laboratory, CAS Centre for Excellence in Quantum Information and Quantum Physics, University of Science and Technology of China, Shanghai 201315, China }
   \author{ Lukasz Dusanowski}%
 \affiliation{Technische Physik and Wilhelm-Conrad-R\"ontgen Research Center for Complex Material Systems, Universit\"at W\"urzburg, Am Hubland, W\"urzburg-97074,	Germany}
 \author{Piotr Andrzej Wronski }
 \affiliation{Technische Physik and Wilhelm-Conrad-R\"ontgen Research Center for Complex Material Systems, Universit\"at W\"urzburg, Am Hubland, W\"urzburg-97074,	Germany}
\author{Chao-Yang Lu}
\affiliation{ National Laboratory for Physical Sciences at Microscale and Department of Modern Physics, University of Science and Technology of China, Shanghai
	Branch, Shanghai 201315, China} 

\affiliation{ CAS-Alibaba Quantum Computing Laboratory, CAS Centre for Excellence in Quantum Information and Quantum Physics, University of Science and Technology of China, Shanghai 201315, China }

\author{ Christian Schneider}%
 
\affiliation{Technische Physik and Wilhelm-Conrad-R\"ontgen Research Center for Complex Material Systems, Universit\"at W\"urzburg, Am Hubland, W\"urzburg-97074,	Germany}

\author{Sven H\"{o}fling}
\email{Sven.Hoefling@physik.uni-wuerzburg.de}
\affiliation{Technische Physik and Wilhelm-Conrad-R\"ontgen Research Center for Complex Material Systems, Universit\"at W\"urzburg, Am Hubland, W\"urzburg-97074,	Germany}
\affiliation{ National Laboratory for Physical Sciences at Microscale and Department of Modern Physics, University of Science and Technology of China, Shanghai
	Branch, Shanghai 201315, China} 
\affiliation{SUPA, School of Physics and Astronomy, University of St. Andrews,
	St. Andrews KY16 9SS, United Kingdom}

\date{\today}
\begin{abstract}
Single photon sources, which are compatible with quantum memories are an important component of quantum networks. In this article, we show optical investigations on isolated GaAs/Al$_{0.25}$Ga$_{0.75}$As quantum dots grown via droplet epitaxy, which emit single photons on resonance with the  Rb-87-D$_2$  line (780 nm). Under continuous wave resonant excitation conditions, we observe bright, clean and narrowband resonance fluorescence emission from such a droplet quantum dot. Furthermore, the second-order correlation measurement clearly demonstrates the single photon emission from this resonantly driven transition. Spectrally resolved resonance fluorescence of a similar quantum dot yields a linewidth as narrow as 660 MHz ($ 2.7~\mu eV $), which corresponds to a coherence time of 0.482 ns. The observed linewidth is  the smallest reported so far for  strain free GaAs quantum dots grown via the droplet method. We believe that this single photon source can be a prime candidate for applications in optical quantum networks.
\end{abstract}

\pacs{Valid PACS appear here}
\maketitle
Long distance quantum communication based on quantum repeater schemes\cite{Sangouard2007} can be realized by a combination of high-performance single photon sources\cite{Aharonovich2016}  and suitable quantum memories.\cite{Lvovsky2009,Fleischhauer2002,Maitre1997,Simon2010a} For this application single photon sources should possess not only high photon fluxes and very narrow linewidth, but also match spectrally with other components of the system\cite{Siyushev2014} such as long-lived quantum memories \cite{Heshami2016} and which may be utilized in  building blocks of quantum repeater nodes.\cite{Rakher2013,Briegel1998} Semiconductor quantum dots (QDs) have been shown to be  good candidates for efficient and highly indistinguishable single photon sources.\cite{Ding2016,Somaschi2016,Unsleber2016a} Moreover, first prototypes of a hybrid semiconductor QD-atomic interface \cite{Akopian2011} have been already realized to demonstrate the principle feasibility of this approach. Coupling of atomic clouds of Rubidium (Rb) with  QDs allowed for demonstration of compact, tunable and spectrally selective delay lines for single photons.\cite{Wildmann2015} Furthermore, the emission wavelength of solid-state GaAs/AlGaAs QDs  were tuned in the spectral range of Rb-87-D$_2$ lines (780 nm)  by introducing strain\cite{Kumar2011,Trotta2016} and even operated under electrical pumping.\cite{Huang2017}

A crucial resource for applications of single photons in  quantum networks\cite{Kok2007} and linear optical computing,\cite{OBrien2007,Pan2012} are high degrees of indistinguishably of the emitted photons. The photons should be identical in color, polarization and their coherence should be Fourier limited.\cite{Schneider2015} However, the quality of single photons emitted from the semiconductor QDs critically depends on excitation conditions. Highest degrees of indistinguishability of single photons have been observed so far under resonant excitation conditions\cite{Ding2016,Somaschi2016,Unsleber2016a} as opposed to  non-resonant excitation. Furthermore, continuous wave (CW) resonance fluorescence provided a paradigm approach to  demonstrate high degrees of indistinguishability and
measure single QD coherence time by probing linewidth of a QD transition, reaching close to lifetime limited values.\cite{Matthiesen2012,Ates2009}

In this work, we focus on a resonance fluorescence measurements of single \\ 
GaAs/Al$_{0.25}$Ga$_{0.75}$As QDs grown by droplet epitaxy. We show that the energy of emitted single photons from selected QDs match the Rb-87-D$_2 $ lines (780 nm).\cite{Siddons2008} Furthermore, the resonance fluorescence measurements yield a spectral linewidth of only 660 MHz corresponding to a coherence time ($ \tau_{coh} $) of 0.482 ns ( using the relation, linewidth = $ \frac{1}{\pi \tau_{coh}}$ ). These measurements strongly outline the feasibility to implement droplet epitaxy grown QDs as single photon sources in quantum information schemes.

Our sample consists of a low-density layer of GaAs/Al$_{0.25}$Ga$_{0.75}$As QDs grown via droplet epitaxy. The   layer of the QDs is embedded in a Schottky-diode structure.  The sample was grown by solid-source molecular beam epitaxy (MBE) on a GaAs(100) semi-insulating a substrate. \cite{Watanabe2000,Langer2014} Ga and Al were supplied by standard effusion cells while  As$_4$ molecules were provided by a valved cracker source. After the growth of a GaAs buffer layer, a 50 nm Si-doped GaAs layers and a 150 nm thick, Si-doped Al$_{0.25}$Ga$_{0.75}$As layer were deposited on the substrate at the temperature of 550 \degree C. The doping concentration was kept constant to be 3$\times $10$ ^{18} $~cm.$^{-3}$ Further, a 20 nm undoped Al$_{0.25}$Ga$_{0.75}$As  tunnel barrier was grown below the QDs at substrate temperature of 350 \degree C and the As valve was closed to reduce  the background As pressure. Moreover,  Ga was deposited at  a flow rate of 0.071 monolayer/s  for 2 minutes and crystallized with As at 3 $ \times  10^{-5}$ Torr beam equivalent pressure. Followed by a  rise of substrate temperature to 400 \degree C, the GaAs QDs were \textit{in-situ} annealed for 10 min to cure  crystal defects. Next, the growth was continued with the AlGaAs  QD capping layer. The overgrowth of QDs with 10 nm  
Al$_{0.25}$Ga $_{0.75}$As was done by migration enhanced epitaxy,\cite{Langer2014} followed by   raising  the substrate temperature to 550 \degree C and 10 periods of a 2 nm AlAs/2 nm Al$_{0.25}$Ga$_{0.75}$As superlattice. The sample growth was completed with 60 nm  Al$_{0.25}$Ga$_{0.75}$As and a 15 nm GaAs cap layer. The areal density of the QDs was found to be  5$\times10^9$~cm$^{-2}$.

For $\mu$ photoluminescence (PL) and resonance fluorescence (RF) measurements the sample was kept in a closed-cycle cryostat at 5K. Non-resonant $ \mu  $PL measurements were done using CW 532 nm laser, while resonantly excited by a 780.076 nm laser. The emitted light from the sample was collected into a spectrometer (1200 lines, grating resolution 0.01 nm) by a 50X objective (NA=0.42). 

\begin{figure}
	\centering
	\includegraphics[scale=0.7]{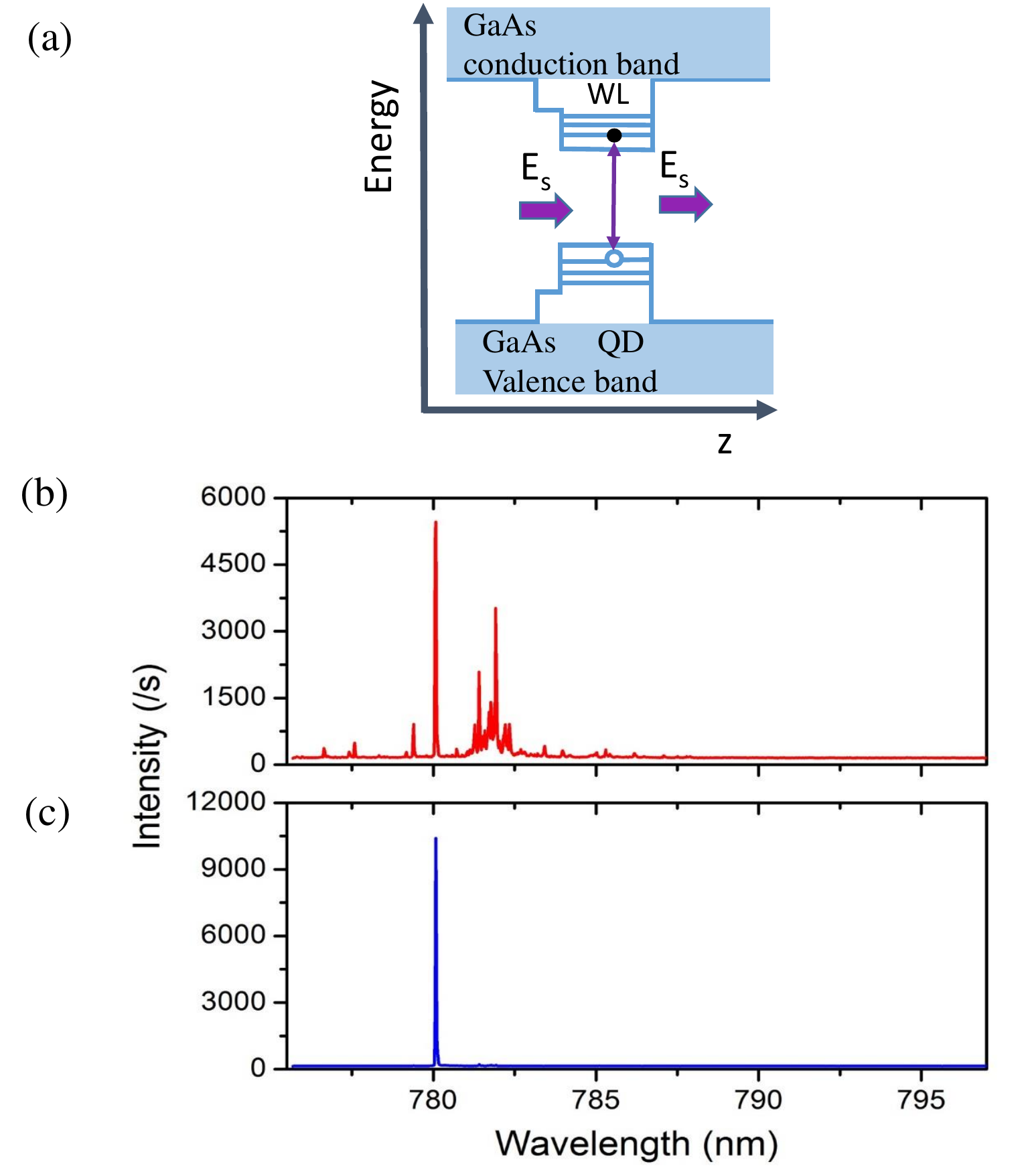}
	\caption[]{\label{NonResonant vs Resonant Fluourocence}  a)  A schematic drawing for  the resonant excitation ($ E_s $) scheme of the  QD. b) Photoluminescence spectrum of a droplet   GaAs/Al$_{0.25}$Ga$_{0.75}$As QD under non-resonant excitation at 532 nm with excitation power of 2 $\mu$W. c) Resonant photoluminescence spectrum, yielding  a single transition line at 780 nm.}
\end{figure}

\begin{figure}
	\centering
	\includegraphics[scale=0.6]{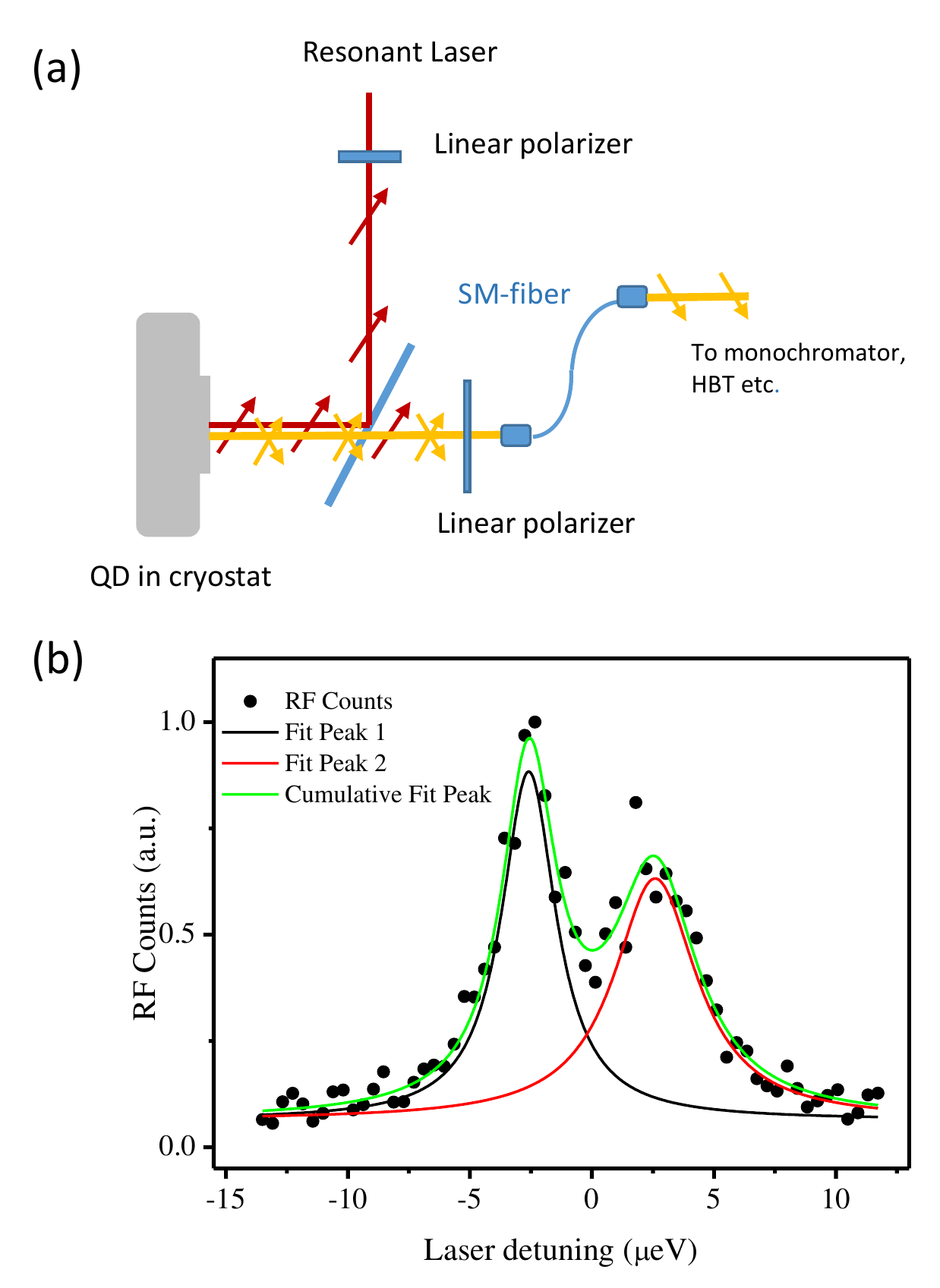}
	\caption[]{\label{Resonance Fluorescence scan} a) A scheme of  cross-polarization resonance fluorescence setup. A laser beam (CW 780.076 nm, resonant to the ground state transition of the QD is passed through a linear polarizer and a  microscope objective to excite the QD. The laser beam reflected from the sample surface is then filtered by a second linear polarizer oriented perpendicular to  the first one. The resonance fluorescence signal emitted from the QD is then collected by the same objective and coupled to a single mode (SM) fiber working as a spatial filter. Finally  the optical signal is introduced to the monochromator equipped with a charged coupled camera detector and Hanbury Brown-Twiss (HBT) interferometer. b) The resonance fluorescence spectrum  recorded by a  laser scan. Solid lines present Lorentzian fits showing two peaks originating from the same QD with a fine structure splitting of about 1.25 GHz. The peak 1 has a linewidth of (0.66 $\pm$ 0.05) GHz and peak 2 has a linewidth of (1.00 $ \pm $ 0.09) GHz.  The signal to noise ratio is~ $  \sim  $11:1. }
\end{figure}
\begin{figure}
	\centering
	\includegraphics[scale=0.6]{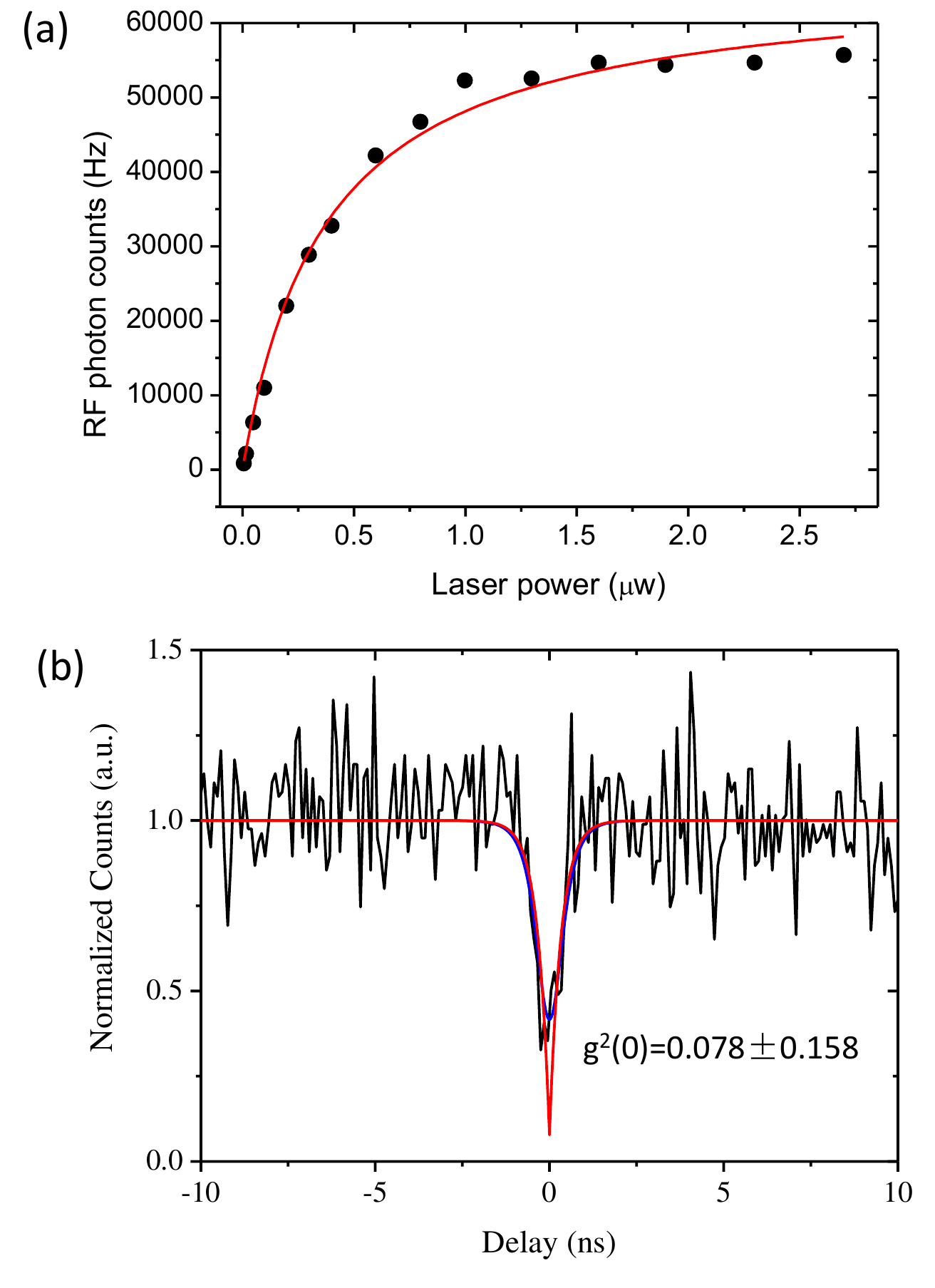}
	\caption[]{\label{g2} (a) Resonance fluorescence photon (RF) counts as a function of increasing  power of  the excitation CW laser. The red line is a fit (using Eq.~1), showing a  saturation of  emission intensity with  increase in excitation power. The best-fit  normalizations power, $P_{n}$ = 0.376	$\mu W$. b) Second order correlation function $ g^2(\tau)$ as a function of time delay between emitted photons. The blue line is the fit to the intensity autocorrelation measurement data without deconvolution. The red line shows deconvolution fit. The best-fit second-order correlation function $g^2(0)$ at zero time delay was found to be $0.078\pm 0.158$.}
\end{figure}
Figure 1b depicts a photoluminescence (PL) spectrum  of  a droplet GaAs/Al$_{0.25}$Ga$_{0.75}$As QDs, which was excited under non-resonant excitation (532 nm). Apart from a distinct QD line at 780 nm, we observe several other QD-attributed emission features, which most likely stem from other QDs excited within laser spot size. The diameter of the laser spot is $ \sim 0.9~ \mu m  $ for 532 nm and  we used 50X magnification objective. Considering the areal density of QDs which is $\sim 5\times\ 10^9~$cm,$^{-2} $ we estimate that about 35 quantum dots might be excited. In order to single out the emission line of interest, we resonantly drive the QD with the fundamental transition at 780 nm with a narrow-band CW laser. As shown in the spectrum in Fig.~1(c), this method clearly yields a purely monochromatic emission by selectively driving the ground state of the QD. The excitation scheme  and energy levels  for a resonant excitation  is shown in  Fig.~1(a).

More importantly, the resonance fluorescence spectrum shows that the emitter of interest can be coherently driven at a frequency which overlaps with the   Rb-87-D$_2 $ lines (780 nm). Further, it is also seen that  the emitter under resonance fluorescence is twice brighter than  the one excited under non-resonant scheme under similar excitation power (2 $\mu$W).

Our method to acquire a pure, nearly background-free resonant emission spectrum from a solid-state quantum emitter is shown in Fig.~2(a). The driving laser is polarized by a combination of a linear polarizer and a half-wave plate, and is focused on the sample surface by a high NA objective. The resonance fluorescence signal from the QD is then collected by the same objective.  A second  perpendicular linear polarizer filters out the laser signal. Spatial filtering is achieved by focusing the collected RF signal into a single mode fiber, before the signal is dispersed in a monochromator and recorded on a CCD camera. In order to accurately deduce the linewidth of our droplet QDs, we performed resonant laser scans under CW excitation [Fig.~2(b)]. 
RF counts were recorded as a function of the laser frequency. There, we chose a pump power of  0.1 $\mu$W, well below the saturation of the QD transition  and an integration time of 20 s. Two resonance fluorescence peaks evolve in the spectrum, which correspond to the fine structure splitting (FSS) of 1.25 GHz of the QD emitter. The FSS is relatively small, due to a greater degree of  structural symmetry\cite{Abbarchi2008,Mano2010} in droplet quantum dots than a Stranski-Krastanov QD.\cite{Huo2013,Huber2017,BassoBasset2018} For example, Huo \textit{et al.}\cite{Huo2013} reported FSS values ranging from 2.4 GHz  to 0.85 GHz.

 In order to confirm the quantum nature of our narrow-band emission feature, we performed power dependent measurements shown in Fig.~3(a). The excitation power dependent PL intensity (Fig.~3a) is fitted with the equation for a two-level system, \cite{Kumar2015a}  
 \begin{equation}
 I = I_{sat} (1/(1+P_n/P_{exc}))~, 
 \end{equation}
  where, $ I_{sat} $ is  the saturation intensity, $P_{exc}$ is  the excitation power and  $ P_{n} $ is  the normalization power.
 A second-order correlation function for a two-level system is given by\cite{He2016a} 
    \begin{equation}
     g^{2} (\tau) =  1 - a \times e^{ -\frac{\abs{ \tau - \tau_0}}{\tau_{c_1}}} , 
  \end{equation}
     here,  $\tau_{c_1}$  accounts for  the exciton relaxation and decay rates of a two-level system and a is a fitting parameter. $\tau$,  $\tau_{0}$ are  the time delay and  the time delay offset, respectively.

 The finite-time resolution of our single-photon detector (350 ps) is accounted for by introducing a Gaussian response function $ G_{d} $  given by\cite{He2016a}
\begin{equation}
  G_{d} = \frac{1} {\sigma \sqrt{\pi}}  e^{-\frac{1}{2} (\frac{\tau-\tau_0}{\sigma})^2}~.
\end{equation}
Here, $ 2 \sqrt{2 ln(2 \sigma)} $ represent 350 ps timing resolution of our avalanche photo diode.
Hence,  the convoluted fitting function,  $g_{real}^{2}(\tau)$ is  given by
\begin{equation}\label{key}
 g_{real}^{2} (\tau)=  G_{d}\otimes  g^{2} (\tau)~.
\end{equation}

Here, $ \otimes $ represent  the convolution operator. In Fig.~3 (b)  $g_{real}^{2}(\tau)$  was used to fit the data to obtain the value of  the second-order correlation  function at zero delay,   $ g^{2} (0)  $ to be $0.078\pm 0.158$ as shown in Fig.~3(b). The non-zero $ g^{2}(0)$ could be  due to  presence of a non-filtered  laser background.

The Schottky diode geometry used in our study enables us to charge tune the emission and spectrally fine tune the emission line of quantized energy levels of the QDs \cite{Langer2014}. We observed resonance fluorescence in both contacted and non-electrically-contacted geometry, and about 50 $ \% $ of the investigated QDs showed resonance fluorescence. We note that in contrast to investigations on high quality InGaAs quantum dots in high quality micropillar cavities,\cite{Ding2016, Wang2016c} the application of additional weak 532 nm laser light on the investigated GaAs/AlGaAs turned out to improve the QD properties similar to other studies, for instance, Gazzano \textit{et al.}\cite{Gazzano2013} The measured linewidths of the doublets in the RF spectrum (660 MHz and 1 GHz) are larger than the   Fourier-transform limit of 488 MHz for the  exciton relaxation time of 330 ps. This difference  can be attributed to an inhomogeneous broadening due to charge noise via the d.c. Stark shift. \cite{Kuhlmann2015} 

In summary, we report a solid state semiconductor single photon device based on isolated  charge-tunable  droplet Al$_{0.25}$Ga$_{0.75}$As QDs. Our resonance fluorescence measurement yield important information about linewidth and fine structure splitting. Most importantly the QD is shown to emit single photons at 780 nm in resonance with Rb-87-D$ _2 $ lines (780 nm).  Continuous wave laser power dependent fluorescence measurement demonstrate two-level behavior of probed emitter. We suggest, that the  QD device may find application in quantum communication devices.

This project has received funding from the European Union’s Horizon 2020 research and innovation programme under the Marie Skłodowska-Curie grant agreement No 721394. We furthermore gratefully acknowledge support by the State of Bavaria.
%


\begin{thebibliography}{36}%
	\makeatletter
	\providecommand \@ifxundefined [1]{%
		\@ifx{#1\undefined}
	}%
	\providecommand \@ifnum [1]{%
		\ifnum #1\expandafter \@firstoftwo
		\else \expandafter \@secondoftwo
		\fi
	}%
	\providecommand \@ifx [1]{%
		\ifx #1\expandafter \@firstoftwo
		\else \expandafter \@secondoftwo
		\fi
	}%
	\providecommand \natexlab [1]{#1}%
	\providecommand \enquote  [1]{``#1''}%
	\providecommand \bibnamefont  [1]{#1}%
	\providecommand \bibfnamefont [1]{#1}%
	\providecommand \citenamefont [1]{#1}%
	\providecommand \href@noop [0]{\@secondoftwo}%
	\providecommand \href [0]{\begingroup \@sanitize@url \@href}%
	\providecommand \@href[1]{\@@startlink{#1}\@@href}%
	\providecommand \@@href[1]{\endgroup#1\@@endlink}%
	\providecommand \@sanitize@url [0]{\catcode `\\12\catcode `\$12\catcode
		`\&12\catcode `\#12\catcode `\^12\catcode `\_12\catcode `\%12\relax}%
	\providecommand \@@startlink[1]{}%
	\providecommand \@@endlink[0]{}%
	\providecommand \url  [0]{\begingroup\@sanitize@url \@url }%
	\providecommand \@url [1]{\endgroup\@href {#1}{\urlprefix }}%
	\providecommand \urlprefix  [0]{URL }%
	\providecommand \Eprint [0]{\href }%
	\providecommand \doibase [0]{http://dx.doi.org/}%
	\providecommand \selectlanguage [0]{\@gobble}%
	\providecommand \bibinfo  [0]{\@secondoftwo}%
	\providecommand \bibfield  [0]{\@secondoftwo}%
	\providecommand \translation [1]{[#1]}%
	\providecommand \BibitemOpen [0]{}%
	\providecommand \bibitemStop [0]{}%
	\providecommand \bibitemNoStop [0]{.\EOS\space}%
	\providecommand \EOS [0]{\spacefactor3000\relax}%
	\providecommand \BibitemShut  [1]{\csname bibitem#1\endcsname}%
	\let\auto@bib@innerbib\@empty
	\bibitem [{\citenamefont {Sangouard}\ \emph {et~al.}(2007)\citenamefont
		{Sangouard}, \citenamefont {Simon}, \citenamefont {Minar}, \citenamefont
		{Zbinden}, \citenamefont {de~Riedmatten},\ and\ \citenamefont
		{Gisin}}]{Sangouard2007}%
	\BibitemOpen
	\bibfield  {author} {\bibinfo {author} {\bibfnamefont {N.}~\bibnamefont
			{Sangouard}}, \bibinfo {author} {\bibfnamefont {C.}~\bibnamefont {Simon}},
		\bibinfo {author} {\bibfnamefont {J.}~\bibnamefont {Minar}}, \bibinfo
		{author} {\bibfnamefont {H.}~\bibnamefont {Zbinden}}, \bibinfo {author}
		{\bibfnamefont {H.}~\bibnamefont {de~Riedmatten}}, \ and\ \bibinfo {author}
		{\bibfnamefont {N.}~\bibnamefont {Gisin}},\ }\href {\doibase
		10.1103/PhysRevA.76.050301} {\bibfield  {journal} {\bibinfo  {journal} {Phys.
				Rev. A}\ }\textbf {\bibinfo {volume} {76}},\ \bibinfo {pages} {050301}
		(\bibinfo {year} {2007})}\BibitemShut {NoStop}%
	\bibitem [{\citenamefont {Aharonovich}\ \emph {et~al.}(2016)\citenamefont
		{Aharonovich}, \citenamefont {Englund},\ and\ \citenamefont
		{Toth}}]{Aharonovich2016}%
	\BibitemOpen
	\bibfield  {author} {\bibinfo {author} {\bibfnamefont {I.}~\bibnamefont
			{Aharonovich}}, \bibinfo {author} {\bibfnamefont {D.}~\bibnamefont
			{Englund}}, \ and\ \bibinfo {author} {\bibfnamefont {M.}~\bibnamefont
			{Toth}},\ }\href {\doibase 10.1038/NPHOTON.2016.186} {\bibfield  {journal}
		{\bibinfo  {journal} {Nat. Photonics}\ }\textbf {\bibinfo {volume} {10}},\
		\bibinfo {pages} {631} (\bibinfo {year} {2016})},\ \bibinfo {note}
	{wOS:000384951900008}\BibitemShut {NoStop}%
	\bibitem [{\citenamefont {Lvovsky}\ \emph {et~al.}(2009)\citenamefont
		{Lvovsky}, \citenamefont {Sanders},\ and\ \citenamefont
		{Tittel}}]{Lvovsky2009}%
	\BibitemOpen
	\bibfield  {author} {\bibinfo {author} {\bibfnamefont {A.~I.}\ \bibnamefont
			{Lvovsky}}, \bibinfo {author} {\bibfnamefont {B.~C.}\ \bibnamefont
			{Sanders}}, \ and\ \bibinfo {author} {\bibfnamefont {W.}~\bibnamefont
			{Tittel}},\ }\href {\doibase 10.1038/nphoton.2009.231} {\bibfield  {journal}
		{\bibinfo  {journal} {Nat. Photonics}\ }\textbf {\bibinfo {volume} {3}},\
		\bibinfo {pages} {706} (\bibinfo {year} {2009})}\BibitemShut {NoStop}%
	\bibitem [{\citenamefont {Fleischhauer}\ and\ \citenamefont
		{Lukin}(2002)}]{Fleischhauer2002}%
	\BibitemOpen
	\bibfield  {author} {\bibinfo {author} {\bibfnamefont {M.}~\bibnamefont
			{Fleischhauer}}\ and\ \bibinfo {author} {\bibfnamefont {M.~D.}\ \bibnamefont
			{Lukin}},\ }\href {\doibase 10.1103/PhysRevA.65.022314} {\bibfield  {journal}
		{\bibinfo  {journal} {Phys. Rev. A}\ }\textbf {\bibinfo {volume} {65}},\
		\bibinfo {pages} {022314} (\bibinfo {year} {2002})}\BibitemShut {NoStop}%
	\bibitem [{\citenamefont {Maitre}\ \emph {et~al.}(1997)\citenamefont {Maitre},
		\citenamefont {Hagley}, \citenamefont {Nogues}, \citenamefont {Wunderlich},
		\citenamefont {Goy}, \citenamefont {Brune}, \citenamefont {Raimond},\ and\
		\citenamefont {Haroche}}]{Maitre1997}%
	\BibitemOpen
	\bibfield  {author} {\bibinfo {author} {\bibfnamefont {X.}~\bibnamefont
			{Maitre}}, \bibinfo {author} {\bibfnamefont {E.}~\bibnamefont {Hagley}},
		\bibinfo {author} {\bibfnamefont {G.}~\bibnamefont {Nogues}}, \bibinfo
		{author} {\bibfnamefont {C.}~\bibnamefont {Wunderlich}}, \bibinfo {author}
		{\bibfnamefont {P.}~\bibnamefont {Goy}}, \bibinfo {author} {\bibfnamefont
			{M.}~\bibnamefont {Brune}}, \bibinfo {author} {\bibfnamefont {J.~M.}\
			\bibnamefont {Raimond}}, \ and\ \bibinfo {author} {\bibfnamefont
			{S.}~\bibnamefont {Haroche}},\ }\href {\doibase 10.1103/PhysRevLett.79.769}
	{\bibfield  {journal} {\bibinfo  {journal} {Phys. Rev. Lett.}\ }\textbf
		{\bibinfo {volume} {79}},\ \bibinfo {pages} {769} (\bibinfo {year}
		{1997})}\BibitemShut {NoStop}%
	\bibitem [{\citenamefont {Simon}\ \emph {et~al.}(2010)\citenamefont {Simon},
		\citenamefont {Afzelius}, \citenamefont {Appel}, \citenamefont {de~la
			Giroday}, \citenamefont {Dewhurst}, \citenamefont {Gisin}, \citenamefont
		{Hu}, \citenamefont {Jelezko}, \citenamefont {Kroll}, \citenamefont {Muller},
		\citenamefont {Nunn}, \citenamefont {Polzik}, \citenamefont {Rarity},
		\citenamefont {De~Riedmatten}, \citenamefont {Rosenfeld}, \citenamefont
		{Shields}, \citenamefont {Skold}, \citenamefont {Stevenson}, \citenamefont
		{Thew}, \citenamefont {Walmsley}, \citenamefont {Weber}, \citenamefont
		{Weinfurter}, \citenamefont {Wrachtrup},\ and\ \citenamefont
		{Young}}]{Simon2010a}%
	\BibitemOpen
	\bibfield  {author} {\bibinfo {author} {\bibfnamefont {C.}~\bibnamefont
			{Simon}}, \bibinfo {author} {\bibfnamefont {M.}~\bibnamefont {Afzelius}},
		\bibinfo {author} {\bibfnamefont {J.}~\bibnamefont {Appel}}, \bibinfo
		{author} {\bibfnamefont {A.~B.}\ \bibnamefont {de~la Giroday}}, \bibinfo
		{author} {\bibfnamefont {S.~J.}\ \bibnamefont {Dewhurst}}, \bibinfo {author}
		{\bibfnamefont {N.}~\bibnamefont {Gisin}}, \bibinfo {author} {\bibfnamefont
			{C.~Y.}\ \bibnamefont {Hu}}, \bibinfo {author} {\bibfnamefont
			{F.}~\bibnamefont {Jelezko}}, \bibinfo {author} {\bibfnamefont
			{S.}~\bibnamefont {Kroll}}, \bibinfo {author} {\bibfnamefont {J.~H.}\
			\bibnamefont {Muller}}, \bibinfo {author} {\bibfnamefont {J.}~\bibnamefont
			{Nunn}}, \bibinfo {author} {\bibfnamefont {E.~S.}\ \bibnamefont {Polzik}},
		\bibinfo {author} {\bibfnamefont {J.~G.}\ \bibnamefont {Rarity}}, \bibinfo
		{author} {\bibfnamefont {H.}~\bibnamefont {De~Riedmatten}}, \bibinfo {author}
		{\bibfnamefont {W.}~\bibnamefont {Rosenfeld}}, \bibinfo {author}
		{\bibfnamefont {A.~J.}\ \bibnamefont {Shields}}, \bibinfo {author}
		{\bibfnamefont {N.}~\bibnamefont {Skold}}, \bibinfo {author} {\bibfnamefont
			{R.~M.}\ \bibnamefont {Stevenson}}, \bibinfo {author} {\bibfnamefont
			{R.}~\bibnamefont {Thew}}, \bibinfo {author} {\bibfnamefont {I.~A.}\
			\bibnamefont {Walmsley}}, \bibinfo {author} {\bibfnamefont {M.~C.}\
			\bibnamefont {Weber}}, \bibinfo {author} {\bibfnamefont {H.}~\bibnamefont
			{Weinfurter}}, \bibinfo {author} {\bibfnamefont {J.}~\bibnamefont
			{Wrachtrup}}, \ and\ \bibinfo {author} {\bibfnamefont {R.~J.}\ \bibnamefont
			{Young}},\ }\href {\doibase 10.1140/epjd/e2010-00103-y} {\bibfield  {journal}
		{\bibinfo  {journal} {Eur. Phys. J. D}\ }\textbf {\bibinfo {volume} {58}},\
		\bibinfo {pages} {1} (\bibinfo {year} {2010})}\BibitemShut {NoStop}%
	\bibitem [{\citenamefont {Siyushev}\ \emph {et~al.}(2014)\citenamefont
		{Siyushev}, \citenamefont {Stein}, \citenamefont {Wrachtrup},\ and\
		\citenamefont {Gerhardt}}]{Siyushev2014}%
	\BibitemOpen
	\bibfield  {author} {\bibinfo {author} {\bibfnamefont {P.}~\bibnamefont
			{Siyushev}}, \bibinfo {author} {\bibfnamefont {G.}~\bibnamefont {Stein}},
		\bibinfo {author} {\bibfnamefont {J.}~\bibnamefont {Wrachtrup}}, \ and\
		\bibinfo {author} {\bibfnamefont {I.}~\bibnamefont {Gerhardt}},\ }\href
	{\doibase 10.1038/nature13191} {\bibfield  {journal} {\bibinfo  {journal}
			{Nature}\ }\textbf {\bibinfo {volume} {509}},\ \bibinfo {pages} {66}
		(\bibinfo {year} {2014})}\BibitemShut {NoStop}%
	\bibitem [{\citenamefont {Heshami}\ \emph {et~al.}(2016)\citenamefont
		{Heshami}, \citenamefont {England}, \citenamefont {Humphreys}, \citenamefont
		{Bustard}, \citenamefont {Acosta}, \citenamefont {Nunn},\ and\ \citenamefont
		{Sussman}}]{Heshami2016}%
	\BibitemOpen
	\bibfield  {author} {\bibinfo {author} {\bibfnamefont {K.}~\bibnamefont
			{Heshami}}, \bibinfo {author} {\bibfnamefont {D.~G.}\ \bibnamefont
			{England}}, \bibinfo {author} {\bibfnamefont {P.~C.}\ \bibnamefont
			{Humphreys}}, \bibinfo {author} {\bibfnamefont {P.~J.}\ \bibnamefont
			{Bustard}}, \bibinfo {author} {\bibfnamefont {V.~M.}\ \bibnamefont {Acosta}},
		\bibinfo {author} {\bibfnamefont {J.}~\bibnamefont {Nunn}}, \ and\ \bibinfo
		{author} {\bibfnamefont {B.~J.}\ \bibnamefont {Sussman}},\ }\href {\doibase
		10.1080/09500340.2016.1148212} {\bibfield  {journal} {\bibinfo  {journal} {J.
				Mod. Opt.}\ }\textbf {\bibinfo {volume} {63}},\ \bibinfo {pages} {2005}
		(\bibinfo {year} {2016})}\BibitemShut {NoStop}%
	\bibitem [{\citenamefont {Rakher}\ \emph {et~al.}(2013)\citenamefont {Rakher},
		\citenamefont {Warburton},\ and\ \citenamefont {Treutlein}}]{Rakher2013}%
	\BibitemOpen
	\bibfield  {author} {\bibinfo {author} {\bibfnamefont {M.~T.}\ \bibnamefont
			{Rakher}}, \bibinfo {author} {\bibfnamefont {R.~J.}\ \bibnamefont
			{Warburton}}, \ and\ \bibinfo {author} {\bibfnamefont {P.}~\bibnamefont
			{Treutlein}},\ }\href {\doibase 10.1103/PhysRevA.88.053834} {\bibfield
		{journal} {\bibinfo  {journal} {Phys. Rev. A}\ }\textbf {\bibinfo {volume}
			{88}},\ \bibinfo {pages} {053834} (\bibinfo {year} {2013})}\BibitemShut
	{NoStop}%
	\bibitem [{\citenamefont {Briegel}\ \emph {et~al.}(1998)\citenamefont
		{Briegel}, \citenamefont {Dur}, \citenamefont {Cirac},\ and\ \citenamefont
		{Zoller}}]{Briegel1998}%
	\BibitemOpen
	\bibfield  {author} {\bibinfo {author} {\bibfnamefont {H.~J.}\ \bibnamefont
			{Briegel}}, \bibinfo {author} {\bibfnamefont {W.}~\bibnamefont {Dur}},
		\bibinfo {author} {\bibfnamefont {J.~I.}\ \bibnamefont {Cirac}}, \ and\
		\bibinfo {author} {\bibfnamefont {P.}~\bibnamefont {Zoller}},\ }\href
	{\doibase 10.1103/PhysRevLett.81.5932} {\bibfield  {journal} {\bibinfo
			{journal} {Phys. Rev. Lett.}\ }\textbf {\bibinfo {volume} {81}},\ \bibinfo
		{pages} {5932} (\bibinfo {year} {1998})}\BibitemShut {NoStop}%
	\bibitem [{\citenamefont {Ding}\ \emph {et~al.}(2016)\citenamefont {Ding},
		\citenamefont {He}, \citenamefont {Duan}, \citenamefont {Gregersen},
		\citenamefont {Chen}, \citenamefont {Unsleber}, \citenamefont {Maier},
		\citenamefont {Schneider}, \citenamefont {Kamp}, \citenamefont {H\"ofling},
		\citenamefont {Lu},\ and\ \citenamefont {Pan}}]{Ding2016}%
	\BibitemOpen
	\bibfield  {author} {\bibinfo {author} {\bibfnamefont {X.}~\bibnamefont
			{Ding}}, \bibinfo {author} {\bibfnamefont {Y.}~\bibnamefont {He}}, \bibinfo
		{author} {\bibfnamefont {Z.-C.}\ \bibnamefont {Duan}}, \bibinfo {author}
		{\bibfnamefont {N.}~\bibnamefont {Gregersen}}, \bibinfo {author}
		{\bibfnamefont {M.-C.}\ \bibnamefont {Chen}}, \bibinfo {author}
		{\bibfnamefont {S.}~\bibnamefont {Unsleber}}, \bibinfo {author}
		{\bibfnamefont {S.}~\bibnamefont {Maier}}, \bibinfo {author} {\bibfnamefont
			{C.}~\bibnamefont {Schneider}}, \bibinfo {author} {\bibfnamefont
			{M.}~\bibnamefont {Kamp}}, \bibinfo {author} {\bibfnamefont {S.}~\bibnamefont
			{H\"ofling}}, \bibinfo {author} {\bibfnamefont {C.-Y.}\ \bibnamefont {Lu}}, \
		and\ \bibinfo {author} {\bibfnamefont {J.-W.}\ \bibnamefont {Pan}},\ }\href
	{\doibase 10.1103/PhysRevLett.116.020401} {\bibfield  {journal} {\bibinfo
			{journal} {Phys. Rev. Lett.}\ }\textbf {\bibinfo {volume} {116}},\ \bibinfo
		{pages} {020401} (\bibinfo {year} {2016})}\BibitemShut {NoStop}%
	\bibitem [{\citenamefont {Somaschi}\ \emph {et~al.}(2016)\citenamefont
		{Somaschi}, \citenamefont {Giesz}, \citenamefont {Santis}, \citenamefont
		{Loredo}, \citenamefont {Almeida}, \citenamefont {Hornecker}, \citenamefont
		{Portalupi}, \citenamefont {Grange}, \citenamefont {Ant{\'{o}}n},
		\citenamefont {Demory}, \citenamefont {G{\'{o}}mez}, \citenamefont {Sagnes},
		\citenamefont {Lanzillotti-Kimura}, \citenamefont {Lema{\'{\i}}tre},
		\citenamefont {Auffeves}, \citenamefont {White}, \citenamefont {Lanco},\ and\
		\citenamefont {Senellart}}]{Somaschi2016}%
	\BibitemOpen
	\bibfield  {author} {\bibinfo {author} {\bibfnamefont {N.}~\bibnamefont
			{Somaschi}}, \bibinfo {author} {\bibfnamefont {V.}~\bibnamefont {Giesz}},
		\bibinfo {author} {\bibfnamefont {L.~D.}\ \bibnamefont {Santis}}, \bibinfo
		{author} {\bibfnamefont {J.~C.}\ \bibnamefont {Loredo}}, \bibinfo {author}
		{\bibfnamefont {M.~P.}\ \bibnamefont {Almeida}}, \bibinfo {author}
		{\bibfnamefont {G.}~\bibnamefont {Hornecker}}, \bibinfo {author}
		{\bibfnamefont {S.~L.}\ \bibnamefont {Portalupi}}, \bibinfo {author}
		{\bibfnamefont {T.}~\bibnamefont {Grange}}, \bibinfo {author} {\bibfnamefont
			{C.}~\bibnamefont {Ant{\'{o}}n}}, \bibinfo {author} {\bibfnamefont
			{J.}~\bibnamefont {Demory}}, \bibinfo {author} {\bibfnamefont
			{C.}~\bibnamefont {G{\'{o}}mez}}, \bibinfo {author} {\bibfnamefont
			{I.}~\bibnamefont {Sagnes}}, \bibinfo {author} {\bibfnamefont {N.~D.}\
			\bibnamefont {Lanzillotti-Kimura}}, \bibinfo {author} {\bibfnamefont
			{A.}~\bibnamefont {Lema{\'{\i}}tre}}, \bibinfo {author} {\bibfnamefont
			{A.}~\bibnamefont {Auffeves}}, \bibinfo {author} {\bibfnamefont {A.~G.}\
			\bibnamefont {White}}, \bibinfo {author} {\bibfnamefont {L.}~\bibnamefont
			{Lanco}}, \ and\ \bibinfo {author} {\bibfnamefont {P.}~\bibnamefont
			{Senellart}},\ }\href {\doibase 10.1038/nphoton.2016.23} {\bibfield
		{journal} {\bibinfo  {journal} {Nat. Photonics}\ }\textbf {\bibinfo {volume}
			{10}},\ \bibinfo {pages} {340} (\bibinfo {year} {2016})}\BibitemShut
	{NoStop}%
	\bibitem [{\citenamefont {Unsleber}\ \emph {et~al.}(2016)\citenamefont
		{Unsleber}, \citenamefont {He}, \citenamefont {Gerhardt}, \citenamefont
		{Maier}, \citenamefont {Lu}, \citenamefont {Pan}, \citenamefont {Gregersen},
		\citenamefont {Kamp}, \citenamefont {Schneider},\ and\ \citenamefont
		{Höfling}}]{Unsleber2016a}%
	\BibitemOpen
	\bibfield  {author} {\bibinfo {author} {\bibfnamefont {S.}~\bibnamefont
			{Unsleber}}, \bibinfo {author} {\bibfnamefont {Y.-M.}\ \bibnamefont {He}},
		\bibinfo {author} {\bibfnamefont {S.}~\bibnamefont {Gerhardt}}, \bibinfo
		{author} {\bibfnamefont {S.}~\bibnamefont {Maier}}, \bibinfo {author}
		{\bibfnamefont {C.-Y.}\ \bibnamefont {Lu}}, \bibinfo {author} {\bibfnamefont
			{J.-W.}\ \bibnamefont {Pan}}, \bibinfo {author} {\bibfnamefont
			{N.}~\bibnamefont {Gregersen}}, \bibinfo {author} {\bibfnamefont
			{M.}~\bibnamefont {Kamp}}, \bibinfo {author} {\bibfnamefont {C.}~\bibnamefont
			{Schneider}}, \ and\ \bibinfo {author} {\bibfnamefont {S.}~\bibnamefont
			{Höfling}},\ }\href {\doibase 10.1364/oe.24.008539} {\bibfield  {journal}
		{\bibinfo  {journal} {Opt. Express}\ }\textbf {\bibinfo {volume} {24}},\
		\bibinfo {pages} {8539} (\bibinfo {year} {2016})}\BibitemShut {NoStop}%
	\bibitem [{\citenamefont {Akopian}\ \emph {et~al.}(2011)\citenamefont
		{Akopian}, \citenamefont {Wang}, \citenamefont {Rastelli}, \citenamefont
		{Schmidt},\ and\ \citenamefont {Zwiller}}]{Akopian2011}%
	\BibitemOpen
	\bibfield  {author} {\bibinfo {author} {\bibfnamefont {N.}~\bibnamefont
			{Akopian}}, \bibinfo {author} {\bibfnamefont {L.}~\bibnamefont {Wang}},
		\bibinfo {author} {\bibfnamefont {A.}~\bibnamefont {Rastelli}}, \bibinfo
		{author} {\bibfnamefont {O.~G.}\ \bibnamefont {Schmidt}}, \ and\ \bibinfo
		{author} {\bibfnamefont {V.}~\bibnamefont {Zwiller}},\ }\href {\doibase
		10.1038/NPHOTON.2011.16} {\bibfield  {journal} {\bibinfo  {journal} {Nat.
				Photonics}\ }\textbf {\bibinfo {volume} {5}},\ \bibinfo {pages} {230}
		(\bibinfo {year} {2011})}\BibitemShut {NoStop}%
	\bibitem [{\citenamefont {Wildmann}\ \emph {et~al.}(2015)\citenamefont
		{Wildmann}, \citenamefont {Trotta}, \citenamefont {Martin-Sanchez},
		\citenamefont {Zallo}, \citenamefont {O'Steen}, \citenamefont {Schmidt},\
		and\ \citenamefont {Rastelli}}]{Wildmann2015}%
	\BibitemOpen
	\bibfield  {author} {\bibinfo {author} {\bibfnamefont {J.~S.}\ \bibnamefont
			{Wildmann}}, \bibinfo {author} {\bibfnamefont {R.}~\bibnamefont {Trotta}},
		\bibinfo {author} {\bibfnamefont {J.}~\bibnamefont {Martin-Sanchez}},
		\bibinfo {author} {\bibfnamefont {E.}~\bibnamefont {Zallo}}, \bibinfo
		{author} {\bibfnamefont {M.}~\bibnamefont {O'Steen}}, \bibinfo {author}
		{\bibfnamefont {O.~G.}\ \bibnamefont {Schmidt}}, \ and\ \bibinfo {author}
		{\bibfnamefont {A.}~\bibnamefont {Rastelli}},\ }\href {\doibase
		10.1103/PhysRevB.92.235306} {\bibfield  {journal} {\bibinfo  {journal} {Phys.
				Rev. B}\ }\textbf {\bibinfo {volume} {92}},\ \bibinfo {pages} {235306}
		(\bibinfo {year} {2015})}\BibitemShut {NoStop}%
	\bibitem [{\citenamefont {Kumar}\ \emph {et~al.}(2011)\citenamefont {Kumar},
		\citenamefont {Trotta}, \citenamefont {Zallo}, \citenamefont {Plumhof},
		\citenamefont {Atkinson}, \citenamefont {Rastelli},\ and\ \citenamefont
		{Schmidt}}]{Kumar2011}%
	\BibitemOpen
	\bibfield  {author} {\bibinfo {author} {\bibfnamefont {S.}~\bibnamefont
			{Kumar}}, \bibinfo {author} {\bibfnamefont {R.}~\bibnamefont {Trotta}},
		\bibinfo {author} {\bibfnamefont {E.}~\bibnamefont {Zallo}}, \bibinfo
		{author} {\bibfnamefont {J.~D.}\ \bibnamefont {Plumhof}}, \bibinfo {author}
		{\bibfnamefont {P.}~\bibnamefont {Atkinson}}, \bibinfo {author}
		{\bibfnamefont {A.}~\bibnamefont {Rastelli}}, \ and\ \bibinfo {author}
		{\bibfnamefont {O.~G.}\ \bibnamefont {Schmidt}},\ }\href {\doibase
		10.1063/1.3653804} {\bibfield  {journal} {\bibinfo  {journal} {Appl. Phys.
				Lett.}\ }\textbf {\bibinfo {volume} {99}},\ \bibinfo {pages} {161118}
		(\bibinfo {year} {2011})}\BibitemShut {NoStop}%
	\bibitem [{\citenamefont {Trotta}\ \emph {et~al.}(2016)\citenamefont {Trotta},
		\citenamefont {Martin-Sanchez}, \citenamefont {Wildmann}, \citenamefont
		{Piredda}, \citenamefont {Reindl}, \citenamefont {Schimpf}, \citenamefont
		{Zallo}, \citenamefont {Stroj}, \citenamefont {Edlinger},\ and\ \citenamefont
		{Rastelli}}]{Trotta2016}%
	\BibitemOpen
	\bibfield  {author} {\bibinfo {author} {\bibfnamefont {R.}~\bibnamefont
			{Trotta}}, \bibinfo {author} {\bibfnamefont {J.}~\bibnamefont
			{Martin-Sanchez}}, \bibinfo {author} {\bibfnamefont {J.~S.}\ \bibnamefont
			{Wildmann}}, \bibinfo {author} {\bibfnamefont {G.}~\bibnamefont {Piredda}},
		\bibinfo {author} {\bibfnamefont {M.}~\bibnamefont {Reindl}}, \bibinfo
		{author} {\bibfnamefont {C.}~\bibnamefont {Schimpf}}, \bibinfo {author}
		{\bibfnamefont {E.}~\bibnamefont {Zallo}}, \bibinfo {author} {\bibfnamefont
			{S.}~\bibnamefont {Stroj}}, \bibinfo {author} {\bibfnamefont
			{J.}~\bibnamefont {Edlinger}}, \ and\ \bibinfo {author} {\bibfnamefont
			{A.}~\bibnamefont {Rastelli}},\ }\href {\doibase 10.1038/ncomms10375}
	{\bibfield  {journal} {\bibinfo  {journal} {Nat. Commun.}\ }\textbf {\bibinfo
			{volume} {7}},\ \bibinfo {pages} {10375} (\bibinfo {year}
		{2016})}\BibitemShut {NoStop}%
	\bibitem [{\citenamefont {Huang}\ \emph {et~al.}(2017)\citenamefont {Huang},
		\citenamefont {Trotta}, \citenamefont {Huo}, \citenamefont {Lettner},
		\citenamefont {Wildmann}, \citenamefont {Mart{\'{\i}}n-S{\'{a}}nchez},
		\citenamefont {Huber}, \citenamefont {Reindl}, \citenamefont {Zhang},
		\citenamefont {Zallo}, \citenamefont {Schmidt},\ and\ \citenamefont
		{Rastelli}}]{Huang2017}%
	\BibitemOpen
	\bibfield  {author} {\bibinfo {author} {\bibfnamefont {H.}~\bibnamefont
			{Huang}}, \bibinfo {author} {\bibfnamefont {R.}~\bibnamefont {Trotta}},
		\bibinfo {author} {\bibfnamefont {Y.}~\bibnamefont {Huo}}, \bibinfo {author}
		{\bibfnamefont {T.}~\bibnamefont {Lettner}}, \bibinfo {author} {\bibfnamefont
			{J.~S.}\ \bibnamefont {Wildmann}}, \bibinfo {author} {\bibfnamefont
			{J.}~\bibnamefont {Mart{\'{\i}}n-S{\'{a}}nchez}}, \bibinfo {author}
		{\bibfnamefont {D.}~\bibnamefont {Huber}}, \bibinfo {author} {\bibfnamefont
			{M.}~\bibnamefont {Reindl}}, \bibinfo {author} {\bibfnamefont
			{J.}~\bibnamefont {Zhang}}, \bibinfo {author} {\bibfnamefont
			{E.}~\bibnamefont {Zallo}}, \bibinfo {author} {\bibfnamefont {O.~G.}\
			\bibnamefont {Schmidt}}, \ and\ \bibinfo {author} {\bibfnamefont
			{A.}~\bibnamefont {Rastelli}},\ }\href {\doibase
		10.1021/acsphotonics.6b00935} {\bibfield  {journal} {\bibinfo  {journal}
			{{ACS} Photonics}\ }\textbf {\bibinfo {volume} {4}},\ \bibinfo {pages} {868}
		(\bibinfo {year} {2017})}\BibitemShut {NoStop}%
	\bibitem [{\citenamefont {Kok}\ \emph {et~al.}(2007)\citenamefont {Kok},
		\citenamefont {Munro}, \citenamefont {Nemoto}, \citenamefont {Ralph},
		\citenamefont {Dowling},\ and\ \citenamefont {Milburn}}]{Kok2007}%
	\BibitemOpen
	\bibfield  {author} {\bibinfo {author} {\bibfnamefont {P.}~\bibnamefont
			{Kok}}, \bibinfo {author} {\bibfnamefont {W.~J.}\ \bibnamefont {Munro}},
		\bibinfo {author} {\bibfnamefont {K.}~\bibnamefont {Nemoto}}, \bibinfo
		{author} {\bibfnamefont {T.~C.}\ \bibnamefont {Ralph}}, \bibinfo {author}
		{\bibfnamefont {J.~P.}\ \bibnamefont {Dowling}}, \ and\ \bibinfo {author}
		{\bibfnamefont {G.~J.}\ \bibnamefont {Milburn}},\ }\href {\doibase
		10.1103/RevModPhys.79.135} {\bibfield  {journal} {\bibinfo  {journal} {Rev.
				Mod. Phys.}\ }\textbf {\bibinfo {volume} {79}},\ \bibinfo {pages} {135}
		(\bibinfo {year} {2007})}\BibitemShut {NoStop}%
	\bibitem [{\citenamefont {O'Brien}(2007)}]{OBrien2007}%
	\BibitemOpen
	\bibfield  {author} {\bibinfo {author} {\bibfnamefont {J.~L.}\ \bibnamefont
			{O'Brien}},\ }\href {\doibase 10.1126/science.1142892} {\bibfield  {journal}
		{\bibinfo  {journal} {Science}\ }\textbf {\bibinfo {volume} {318}},\ \bibinfo
		{pages} {1567} (\bibinfo {year} {2007})}\BibitemShut {NoStop}%
	\bibitem [{\citenamefont {Pan}\ \emph {et~al.}(2012)\citenamefont {Pan},
		\citenamefont {Chen}, \citenamefont {Lu}, \citenamefont {Weinfurter},
		\citenamefont {Zeilinger},\ and\ \citenamefont {\ifmmode~\dot{Z}\else
			\.{Z}\fi{}ukowski}}]{Pan2012}%
	\BibitemOpen
	\bibfield  {author} {\bibinfo {author} {\bibfnamefont {J.-W.}\ \bibnamefont
			{Pan}}, \bibinfo {author} {\bibfnamefont {Z.-B.}\ \bibnamefont {Chen}},
		\bibinfo {author} {\bibfnamefont {C.-Y.}\ \bibnamefont {Lu}}, \bibinfo
		{author} {\bibfnamefont {H.}~\bibnamefont {Weinfurter}}, \bibinfo {author}
		{\bibfnamefont {A.}~\bibnamefont {Zeilinger}}, \ and\ \bibinfo {author}
		{\bibfnamefont {M.}~\bibnamefont {\ifmmode~\dot{Z}\else \.{Z}\fi{}ukowski}},\
	}\href {\doibase 10.1103/RevModPhys.84.777} {\bibfield  {journal} {\bibinfo
			{journal} {Rev. Mod. Phys.}\ }\textbf {\bibinfo {volume} {84}},\ \bibinfo
		{pages} {777} (\bibinfo {year} {2012})}\BibitemShut {NoStop}%
	\bibitem [{\citenamefont {Schneider}\ \emph {et~al.}(2015)\citenamefont
		{Schneider}, \citenamefont {Gold}, \citenamefont {Lu}, \citenamefont
		{H{\"o}fling}, \citenamefont {Pan},\ and\ \citenamefont
		{Kamp}}]{Schneider2015}%
	\BibitemOpen
	\bibfield  {author} {\bibinfo {author} {\bibfnamefont {C.}~\bibnamefont
			{Schneider}}, \bibinfo {author} {\bibfnamefont {P.}~\bibnamefont {Gold}},
		\bibinfo {author} {\bibfnamefont {C.-Y.}\ \bibnamefont {Lu}}, \bibinfo
		{author} {\bibfnamefont {S.}~\bibnamefont {H{\"o}fling}}, \bibinfo {author}
		{\bibfnamefont {J.-W.}\ \bibnamefont {Pan}}, \ and\ \bibinfo {author}
		{\bibfnamefont {M.}~\bibnamefont {Kamp}},\ }in\ \href@noop {} {\emph
		{\bibinfo {booktitle} {Engineering the Atom-Photon Interaction}}}\ (\bibinfo
	{publisher} {Springer},\ \bibinfo {year} {2015})\ pp.\ \bibinfo {pages}
	{343--361}\BibitemShut {NoStop}%
	\bibitem [{\citenamefont {Matthiesen}\ \emph {et~al.}(2012)\citenamefont
		{Matthiesen}, \citenamefont {Vamivakas},\ and\ \citenamefont
		{Atatuere}}]{Matthiesen2012}%
	\BibitemOpen
	\bibfield  {author} {\bibinfo {author} {\bibfnamefont {C.}~\bibnamefont
			{Matthiesen}}, \bibinfo {author} {\bibfnamefont {A.~N.}\ \bibnamefont
			{Vamivakas}}, \ and\ \bibinfo {author} {\bibfnamefont {M.}~\bibnamefont
			{Atatuere}},\ }\href {\doibase 10.1103/PhysRevLett.108.093602} {\bibfield
		{journal} {\bibinfo  {journal} {Phys. Rev. Lett.}\ }\textbf {\bibinfo
			{volume} {108}},\ \bibinfo {pages} {093602} (\bibinfo {year}
		{2012})}\BibitemShut {NoStop}%
	\bibitem [{\citenamefont {Ates}\ \emph {et~al.}(2009)\citenamefont {Ates},
		\citenamefont {Ulrich}, \citenamefont {Ulhaq}, \citenamefont {Reitzenstein},
		\citenamefont {Loffler}, \citenamefont {Hofling}, \citenamefont {Forchel},\
		and\ \citenamefont {Michler}}]{Ates2009}%
	\BibitemOpen
	\bibfield  {author} {\bibinfo {author} {\bibfnamefont {S.}~\bibnamefont
			{Ates}}, \bibinfo {author} {\bibfnamefont {S.~M.}\ \bibnamefont {Ulrich}},
		\bibinfo {author} {\bibfnamefont {A.}~\bibnamefont {Ulhaq}}, \bibinfo
		{author} {\bibfnamefont {S.}~\bibnamefont {Reitzenstein}}, \bibinfo {author}
		{\bibfnamefont {A.}~\bibnamefont {Loffler}}, \bibinfo {author} {\bibfnamefont
			{S.}~\bibnamefont {Hofling}}, \bibinfo {author} {\bibfnamefont
			{A.}~\bibnamefont {Forchel}}, \ and\ \bibinfo {author} {\bibfnamefont
			{P.}~\bibnamefont {Michler}},\ }\href {\doibase 10.1038/nphoton.2009.215}
	{\bibfield  {journal} {\bibinfo  {journal} {Nat. Photonics}\ }\textbf
		{\bibinfo {volume} {3}},\ \bibinfo {pages} {724} (\bibinfo {year}
		{2009})}\BibitemShut {NoStop}%
	\bibitem [{\citenamefont {Siddons}\ \emph {et~al.}(2008)\citenamefont
		{Siddons}, \citenamefont {Adams}, \citenamefont {Ge},\ and\ \citenamefont
		{Hughes}}]{Siddons2008}%
	\BibitemOpen
	\bibfield  {author} {\bibinfo {author} {\bibfnamefont {P.}~\bibnamefont
			{Siddons}}, \bibinfo {author} {\bibfnamefont {C.~S.}\ \bibnamefont {Adams}},
		\bibinfo {author} {\bibfnamefont {C.}~\bibnamefont {Ge}}, \ and\ \bibinfo
		{author} {\bibfnamefont {I.~G.}\ \bibnamefont {Hughes}},\ }\href {\doibase
		10.1088/0953-4075/41/15/155004} {\bibfield  {journal} {\bibinfo  {journal}
			{Journal of Physics B-atomic Molecular and Optical Physics}\ }\textbf
		{\bibinfo {volume} {41}},\ \bibinfo {pages} {155004} (\bibinfo {year}
		{2008})}\BibitemShut {NoStop}%
	\bibitem [{\citenamefont {Watanabe}\ \emph {et~al.}(2000)\citenamefont
		{Watanabe}, \citenamefont {Koguchi},\ and\ \citenamefont
		{Gotoh}}]{Watanabe2000}%
	\BibitemOpen
	\bibfield  {author} {\bibinfo {author} {\bibfnamefont {K.}~\bibnamefont
			{Watanabe}}, \bibinfo {author} {\bibfnamefont {N.}~\bibnamefont {Koguchi}}, \
		and\ \bibinfo {author} {\bibfnamefont {Y.}~\bibnamefont {Gotoh}},\ }\href
	{\doibase 10.1143/JJAP.39.L79} {\bibfield  {journal} {\bibinfo  {journal}
			{Japanese Journal of Applied Physics Part 2-letters}\ }\textbf {\bibinfo
			{volume} {39}},\ \bibinfo {pages} {L79} (\bibinfo {year} {2000})}\BibitemShut
	{NoStop}%
	\bibitem [{\citenamefont {Langer}\ \emph {et~al.}(2014)\citenamefont {Langer},
		\citenamefont {Plischke}, \citenamefont {Kamp},\ and\ \citenamefont
		{Höfling}}]{Langer2014}%
	\BibitemOpen
	\bibfield  {author} {\bibinfo {author} {\bibfnamefont {F.}~\bibnamefont
			{Langer}}, \bibinfo {author} {\bibfnamefont {D.}~\bibnamefont {Plischke}},
		\bibinfo {author} {\bibfnamefont {M.}~\bibnamefont {Kamp}}, \ and\ \bibinfo
		{author} {\bibfnamefont {S.}~\bibnamefont {Höfling}},\ }\href {\doibase
		10.1063/1.4894372} {\bibfield  {journal} {\bibinfo  {journal} {Appl. Phys.
				Lett.}\ }\textbf {\bibinfo {volume} {105}},\ \bibinfo {pages} {081111}
		(\bibinfo {year} {2014})}\BibitemShut {NoStop}%
	\bibitem [{\citenamefont {Abbarchi}\ \emph {et~al.}(2008)\citenamefont
		{Abbarchi}, \citenamefont {Mastrandrea}, \citenamefont {Kuroda},
		\citenamefont {Mano}, \citenamefont {Sakoda}, \citenamefont {Koguchi},
		\citenamefont {Sanguinetti}, \citenamefont {Vinattieri},\ and\ \citenamefont
		{Gurioli}}]{Abbarchi2008}%
	\BibitemOpen
	\bibfield  {author} {\bibinfo {author} {\bibfnamefont {M.}~\bibnamefont
			{Abbarchi}}, \bibinfo {author} {\bibfnamefont {C.~A.}\ \bibnamefont
			{Mastrandrea}}, \bibinfo {author} {\bibfnamefont {T.}~\bibnamefont {Kuroda}},
		\bibinfo {author} {\bibfnamefont {T.}~\bibnamefont {Mano}}, \bibinfo {author}
		{\bibfnamefont {K.}~\bibnamefont {Sakoda}}, \bibinfo {author} {\bibfnamefont
			{N.}~\bibnamefont {Koguchi}}, \bibinfo {author} {\bibfnamefont
			{S.}~\bibnamefont {Sanguinetti}}, \bibinfo {author} {\bibfnamefont
			{A.}~\bibnamefont {Vinattieri}}, \ and\ \bibinfo {author} {\bibfnamefont
			{M.}~\bibnamefont {Gurioli}},\ }\href {\doibase 10.1103/PhysRevB.78.125321}
	{\bibfield  {journal} {\bibinfo  {journal} {Phys. Rev. B}\ }\textbf {\bibinfo
			{volume} {78}},\ \bibinfo {pages} {125321} (\bibinfo {year}
		{2008})}\BibitemShut {NoStop}%
	\bibitem [{\citenamefont {Mano}\ \emph {et~al.}(2010)\citenamefont {Mano},
		\citenamefont {Abbarchi}, \citenamefont {Kuroda}, \citenamefont {McSkimming},
		\citenamefont {Ohtake}, \citenamefont {Mitsuishi},\ and\ \citenamefont
		{Sakoda}}]{Mano2010}%
	\BibitemOpen
	\bibfield  {author} {\bibinfo {author} {\bibfnamefont {T.}~\bibnamefont
			{Mano}}, \bibinfo {author} {\bibfnamefont {M.}~\bibnamefont {Abbarchi}},
		\bibinfo {author} {\bibfnamefont {T.}~\bibnamefont {Kuroda}}, \bibinfo
		{author} {\bibfnamefont {B.}~\bibnamefont {McSkimming}}, \bibinfo {author}
		{\bibfnamefont {A.}~\bibnamefont {Ohtake}}, \bibinfo {author} {\bibfnamefont
			{K.}~\bibnamefont {Mitsuishi}}, \ and\ \bibinfo {author} {\bibfnamefont
			{K.}~\bibnamefont {Sakoda}},\ }\href {\doibase 10.1143/apex.3.065203}
	{\bibfield  {journal} {\bibinfo  {journal} {Appl. Phys Express}\ }\textbf
		{\bibinfo {volume} {3}},\ \bibinfo {pages} {065203} (\bibinfo {year}
		{2010})}\BibitemShut {NoStop}%
	\bibitem [{\citenamefont {Huo}\ \emph {et~al.}(2013)\citenamefont {Huo},
		\citenamefont {Rastelli},\ and\ \citenamefont {Schmidt}}]{Huo2013}%
	\BibitemOpen
	\bibfield  {author} {\bibinfo {author} {\bibfnamefont {Y.}~\bibnamefont
			{Huo}}, \bibinfo {author} {\bibfnamefont {A.}~\bibnamefont {Rastelli}}, \
		and\ \bibinfo {author} {\bibfnamefont {O.}~\bibnamefont {Schmidt}},\ }\href
	{\doibase 10.1063/1.4802088} {\bibfield  {journal} {\bibinfo  {journal}
			{Appl. Phys. Lett.}\ }\textbf {\bibinfo {volume} {102}},\ \bibinfo {pages}
		{152105} (\bibinfo {year} {2013})}\BibitemShut {NoStop}%
	\bibitem [{\citenamefont {Huber}\ \emph {et~al.}(2017)\citenamefont {Huber},
		\citenamefont {Reindl}, \citenamefont {Huo}, \citenamefont {Huang},
		\citenamefont {Wildmann}, \citenamefont {Schmidt}, \citenamefont {Rastelli},\
		and\ \citenamefont {Trotta}}]{Huber2017}%
	\BibitemOpen
	\bibfield  {author} {\bibinfo {author} {\bibfnamefont {D.}~\bibnamefont
			{Huber}}, \bibinfo {author} {\bibfnamefont {M.}~\bibnamefont {Reindl}},
		\bibinfo {author} {\bibfnamefont {Y.}~\bibnamefont {Huo}}, \bibinfo {author}
		{\bibfnamefont {H.}~\bibnamefont {Huang}}, \bibinfo {author} {\bibfnamefont
			{J.~S.}\ \bibnamefont {Wildmann}}, \bibinfo {author} {\bibfnamefont {O.~G.}\
			\bibnamefont {Schmidt}}, \bibinfo {author} {\bibfnamefont {A.}~\bibnamefont
			{Rastelli}}, \ and\ \bibinfo {author} {\bibfnamefont {R.}~\bibnamefont
			{Trotta}},\ }\href {\doibase 10.1038/ncomms15506} {\bibfield  {journal}
		{\bibinfo  {journal} {Nat. Commun.}\ }\textbf {\bibinfo {volume} {8}},\
		\bibinfo {pages} {15506} (\bibinfo {year} {2017})}\BibitemShut {NoStop}%
	\bibitem [{\citenamefont {Basso~Basset}\ \emph {et~al.}(2018)\citenamefont
		{Basso~Basset}, \citenamefont {Bietti}, \citenamefont {Reindl}, \citenamefont
		{Esposito}, \citenamefont {Fedorov}, \citenamefont {Huber}, \citenamefont
		{Rastelli}, \citenamefont {Bonera}, \citenamefont {Trotta},\ and\
		\citenamefont {Sanguinetti}}]{BassoBasset2018}%
	\BibitemOpen
	\bibfield  {author} {\bibinfo {author} {\bibfnamefont {F.}~\bibnamefont
			{Basso~Basset}}, \bibinfo {author} {\bibfnamefont {S.}~\bibnamefont
			{Bietti}}, \bibinfo {author} {\bibfnamefont {M.}~\bibnamefont {Reindl}},
		\bibinfo {author} {\bibfnamefont {L.}~\bibnamefont {Esposito}}, \bibinfo
		{author} {\bibfnamefont {A.}~\bibnamefont {Fedorov}}, \bibinfo {author}
		{\bibfnamefont {D.}~\bibnamefont {Huber}}, \bibinfo {author} {\bibfnamefont
			{A.}~\bibnamefont {Rastelli}}, \bibinfo {author} {\bibfnamefont
			{E.}~\bibnamefont {Bonera}}, \bibinfo {author} {\bibfnamefont
			{R.}~\bibnamefont {Trotta}}, \ and\ \bibinfo {author} {\bibfnamefont
			{S.}~\bibnamefont {Sanguinetti}},\ }\href {\doibase
		10.1021/acs.nanolett.7b04472} {\bibfield  {journal} {\bibinfo  {journal}
			{Nano Lett.}\ }\textbf {\bibinfo {volume} {18}},\ \bibinfo {pages} {505}
		(\bibinfo {year} {2018})}\BibitemShut {NoStop}%
	\bibitem [{\citenamefont {Kumar}\ \emph {et~al.}(2015)\citenamefont {Kumar},
		\citenamefont {Kaczmarczyk},\ and\ \citenamefont {Gerardot}}]{Kumar2015a}%
	\BibitemOpen
	\bibfield  {author} {\bibinfo {author} {\bibfnamefont {S.}~\bibnamefont
			{Kumar}}, \bibinfo {author} {\bibfnamefont {A.}~\bibnamefont {Kaczmarczyk}},
		\ and\ \bibinfo {author} {\bibfnamefont {B.~D.}\ \bibnamefont {Gerardot}},\
	}\bibfield  {booktitle} {\emph {\bibinfo {booktitle} {Nano Letters}},\ }\href
	{\doibase 10.1021/acs.nanolett.5b03312} {\bibfield  {journal} {\bibinfo
			{journal} {Nano Lett.}\ }\textbf {\bibinfo {volume} {15}},\ \bibinfo {pages}
		{7567} (\bibinfo {year} {2015})}\BibitemShut {NoStop}%
	\bibitem [{\citenamefont {He}\ \emph {et~al.}(2016)\citenamefont {He},
		\citenamefont {Iff}, \citenamefont {Lundt}, \citenamefont {Baumann},
		\citenamefont {Davanco}, \citenamefont {Srinivasan}, \citenamefont
		{Höfling},\ and\ \citenamefont {Schneider}}]{He2016a}%
	\BibitemOpen
	\bibfield  {author} {\bibinfo {author} {\bibfnamefont {Y.-M.}\ \bibnamefont
			{He}}, \bibinfo {author} {\bibfnamefont {O.}~\bibnamefont {Iff}}, \bibinfo
		{author} {\bibfnamefont {N.}~\bibnamefont {Lundt}}, \bibinfo {author}
		{\bibfnamefont {V.}~\bibnamefont {Baumann}}, \bibinfo {author} {\bibfnamefont
			{M.}~\bibnamefont {Davanco}}, \bibinfo {author} {\bibfnamefont
			{K.}~\bibnamefont {Srinivasan}}, \bibinfo {author} {\bibfnamefont
			{S.}~\bibnamefont {Höfling}}, \ and\ \bibinfo {author} {\bibfnamefont
			{C.}~\bibnamefont {Schneider}},\ }\href
	{http://dx.doi.org/10.1038/ncomms13409} {\bibfield  {journal} {\bibinfo
			{journal} {Nat. Commun.}\ }\textbf {\bibinfo {volume} {7}},\ \bibinfo {pages}
		{13409} (\bibinfo {year} {2016})}\BibitemShut {NoStop}%
	\bibitem [{\citenamefont {Gazzano}\ \emph {et~al.}(2013)\citenamefont
		{Gazzano}, \citenamefont {Michaelis~de Vasconcellos}, \citenamefont {Arnold},
		\citenamefont {Nowak}, \citenamefont {Galopin}, \citenamefont {Sagnes},
		\citenamefont {Lanco}, \citenamefont {Lemaître},\ and\ \citenamefont
		{Senellart}}]{Gazzano2013}%
	\BibitemOpen
	\bibfield  {author} {\bibinfo {author} {\bibfnamefont {O.}~\bibnamefont
			{Gazzano}}, \bibinfo {author} {\bibfnamefont {S.}~\bibnamefont {Michaelis~de
				Vasconcellos}}, \bibinfo {author} {\bibfnamefont {C.}~\bibnamefont {Arnold}},
		\bibinfo {author} {\bibfnamefont {A.}~\bibnamefont {Nowak}}, \bibinfo
		{author} {\bibfnamefont {E.}~\bibnamefont {Galopin}}, \bibinfo {author}
		{\bibfnamefont {I.}~\bibnamefont {Sagnes}}, \bibinfo {author} {\bibfnamefont
			{L.}~\bibnamefont {Lanco}}, \bibinfo {author} {\bibfnamefont
			{A.}~\bibnamefont {Lemaître}}, \ and\ \bibinfo {author} {\bibfnamefont
			{P.}~\bibnamefont {Senellart}},\ }\href
	{http://dx.doi.org/10.1038/ncomms2434} {\bibfield  {journal} {\bibinfo
			{journal} {Nat. Commun.}\ }\textbf {\bibinfo {volume} {4}},\ \bibinfo {pages}
		{1425} (\bibinfo {year} {2013})}\BibitemShut {NoStop}%
	\bibitem [{\citenamefont {Wang}\ \emph {et~al.}(2016)\citenamefont {Wang},
		\citenamefont {Duan}, \citenamefont {Li}, \citenamefont {Chen}, \citenamefont
		{Li}, \citenamefont {He}, \citenamefont {Chen}, \citenamefont {He},
		\citenamefont {Ding}, \citenamefont {Peng}, \citenamefont {Schneider},
		\citenamefont {Kamp}, \citenamefont {H\"ofling}, \citenamefont {Lu},\ and\
		\citenamefont {Pan}}]{Wang2016c}%
	\BibitemOpen
	\bibfield  {author} {\bibinfo {author} {\bibfnamefont {H.}~\bibnamefont
			{Wang}}, \bibinfo {author} {\bibfnamefont {Z.-C.}\ \bibnamefont {Duan}},
		\bibinfo {author} {\bibfnamefont {Y.-H.}\ \bibnamefont {Li}}, \bibinfo
		{author} {\bibfnamefont {S.}~\bibnamefont {Chen}}, \bibinfo {author}
		{\bibfnamefont {J.-P.}\ \bibnamefont {Li}}, \bibinfo {author} {\bibfnamefont
			{Y.-M.}\ \bibnamefont {He}}, \bibinfo {author} {\bibfnamefont {M.-C.}\
			\bibnamefont {Chen}}, \bibinfo {author} {\bibfnamefont {Y.}~\bibnamefont
			{He}}, \bibinfo {author} {\bibfnamefont {X.}~\bibnamefont {Ding}}, \bibinfo
		{author} {\bibfnamefont {C.-Z.}\ \bibnamefont {Peng}}, \bibinfo {author}
		{\bibfnamefont {C.}~\bibnamefont {Schneider}}, \bibinfo {author}
		{\bibfnamefont {M.}~\bibnamefont {Kamp}}, \bibinfo {author} {\bibfnamefont
			{S.}~\bibnamefont {H\"ofling}}, \bibinfo {author} {\bibfnamefont {C.-Y.}\
			\bibnamefont {Lu}}, \ and\ \bibinfo {author} {\bibfnamefont {J.-W.}\
			\bibnamefont {Pan}},\ }\href {\doibase 10.1103/PhysRevLett.116.213601}
	{\bibfield  {journal} {\bibinfo  {journal} {Phys. Rev. Lett.}\ }\textbf
		{\bibinfo {volume} {116}},\ \bibinfo {pages} {213601} (\bibinfo {year}
		{2016})}\BibitemShut {NoStop}%
	\bibitem [{\citenamefont {Kuhlmann}\ \emph {et~al.}(2015)\citenamefont
		{Kuhlmann}, \citenamefont {Prechtel}, \citenamefont {Houel}, \citenamefont
		{Ludwig}, \citenamefont {Reuter}, \citenamefont {Wieck},\ and\ \citenamefont
		{Warburton}}]{Kuhlmann2015}%
	\BibitemOpen
	\bibfield  {author} {\bibinfo {author} {\bibfnamefont {A.~V.}\ \bibnamefont
			{Kuhlmann}}, \bibinfo {author} {\bibfnamefont {J.~H.}\ \bibnamefont
			{Prechtel}}, \bibinfo {author} {\bibfnamefont {J.}~\bibnamefont {Houel}},
		\bibinfo {author} {\bibfnamefont {A.}~\bibnamefont {Ludwig}}, \bibinfo
		{author} {\bibfnamefont {D.}~\bibnamefont {Reuter}}, \bibinfo {author}
		{\bibfnamefont {A.~D.}\ \bibnamefont {Wieck}}, \ and\ \bibinfo {author}
		{\bibfnamefont {R.~J.}\ \bibnamefont {Warburton}},\ }\href
	{http://dx.doi.org/10.1038/ncomms9204} {\bibfield  {journal} {\bibinfo
			{journal} {Nat. Commun.}\ }\textbf {\bibinfo {volume} {6}},\  (\bibinfo
		{year} {2015})}\BibitemShut {NoStop}%
\end{thebibliography}
\end{document}